# The Impact of IMSI Catcher Deployments on Cellular Network Security: Challenges and Countermeasures in 4G and 5G Networks

**Karwan Mustafa Kareem**
MSc Advanced Computer Science – United Kingdom
University of Sulaimani, Kurdistan Region, Iraq
Sulaymaniyah, Iraq
e-mail: karwan.kareem@univsul.edu.iq

**Abstract**—IMSI (International Mobile Subscriber Identity) catchers, also known as "Stingrays" or "cell site simulators," are rogue devices that pose a significant threat to cellular network security [1]. IMSI catchers can intercept and manipulate cellular communications, compromising the privacy and security of mobile devices and their users. With the advent of 4G and 5G networks, IMSI catchers have become more sophisticated and pose new challenges to cellular network security [2].

This paper provides an overview of the impact of IMSI catcher deployments on cellular network security in the context of 4G and 5G networks. It discusses the challenges posed by IMSI catchers, including the unauthorized collection of IMSI numbers, interception of communications, and potential misuse of subscriber information. It also highlights the potential consequences of IMSI catcher deployments, including the compromise of user privacy, financial fraud, and unauthorized surveillance.

The paper further reviews the countermeasures that can be employed to mitigate the risks posed by IMSI catchers. These countermeasures include network-based solutions such as signal analysis, encryption, and authentication mechanisms, as well as user-based solutions such as mobile applications and device settings. The paper also discusses the limitations and effectiveness of these countermeasures in the context of 4G and 5G networks.

Finally, the paper identifies research gaps and future directions for enhancing cellular network security against IMSI catchers in the era of 4G and 5G networks. This includes the need for improved encryption algorithms, authentication mechanisms, and detection techniques to effectively detect and prevent IMSI catcher deployments. The paper also emphasizes the importance of regulatory and policy measures to govern the deployment and use of IMSI catchers to protect user privacy and security.

**Keywords**-IMSI catchers, Stingrays, cellular network security, 4G, 5G, challenges, countermeasures, privacy, authentication, encryption, detection.

## I. Introduction

The rapid advancements in mobile communication technologies, particularly with the deployment of 4G and 5G networks, have revolutionized the way we communicate and access information. These networks offer higher data rates, lower latency, and increased connectivity, enabling a wide range of applications and services. However, along with these benefits, there are also concerns about the security and privacy of cellular networks. One significant threat to cellular network security is the deployment of IMSI catchers, also known as "Stingrays" or "cell site simulators." IMSI catchers are rogue devices that can intercept and manipulate cellular communications, compromising the privacy and security of mobile devices and their users [3].

IMSI catchers work by mimicking legitimate cellular base stations and tricking nearby mobile devices into connecting to them. Once connected, the IMSI catcher can intercept and record communications, including voice calls, text messages, and data traffic [4]. They can also collect International Mobile Subscriber Identity (IMSI) numbers, which are unique identifiers associated with mobile devices, allowing the attacker to track and identify individual users [6]. IMSI catchers can be used for various purposes, including surveillance, espionage, financial fraud, and identity theft [5].

The deployment of IMSI catchers poses significant challenges to cellular network security, particularly in the context of 4G and 5G networks [7], [8]. These networks use complex technologies, such as Long-Term Evolution (LTE) and New Radio (NR), which introduce new vulnerabilities that can be exploited by IMSI catchers. For example, IMSI catchers can exploit vulnerabilities in the LTE and NR authentication and encryption mechanisms to intercept and manipulate communications [9]. Moreover, the increasing use of mobile devices and the growing demand for mobile data services make cellular networks more susceptible to IMSI catcher attacks.

To address the threat posed by IMSI catchers, various countermeasures can be employed. These countermeasures include network-based solutions, such as signal analysis, encryption, and authentication mechanisms, as well as user-based solutions, such as mobile applications and device settings [10]. However, these countermeasures also have limitations, and their effectiveness may vary depending on the specific network environment and deployment scenario. Therefore, it is crucial to understand the challenges and limitations of these countermeasures to effectively mitigate the risks posed by IMSI catchers in the era of 4G and 5G networks.

In this paper, we provide an overview of the impact of IMSI catcher deployments on cellular network security in the context of 4G and 5G networks. We discuss the challenges posed by IMSI catchers, including the unauthorized collection of IMSI numbers, interception of communications, and potential misuse of subscriber information. We also highlight the potential consequences of IMSI catcher deployments, including the compromise of user privacy, financial fraud, and unauthorized surveillance [11]. Furthermore, we review the countermeasures that can be employed to mitigate the risks posed by IMSI catchers, including network-based solutions and user-based solutions. We discuss the limitations and effectiveness of these countermeasures in the context of 4G and 5G networks. Finally, we identify research gaps and future directions for enhancing cellular network security against IMSI catchers.

## II. Literature Review

Dabrowski et al. (2014) discuss the growing threat posed by IMSI-catchers, which are devices used for intercepting and tracking mobile phone communications by spoofing cell towers. The authors highlight that while these devices are mainly used by law enforcement agencies to track suspects, they can also be utilized by criminals for eavesdropping and stealing sensitive information. To counter this threat, the paper suggests the use of "IMSI-catcher-catchers," which are devices that can detect the presence of IMSI-catchers and alert users to their presence. The authors discuss the development of a prototype IMSI-catcher-catcher and its effectiveness in detecting IMSI-catchers in real-world scenarios.

As a whole, this paper emphasizes the importance of mobile phone security at a time when surveillance technologies, such as IMSI-catchers, are becoming increasingly common. The authors argue that the development of IMSI-catcher-catchers represents a crucial step toward protecting individual privacy and combating mobile phone surveillance. This paper contributes to the literature on mobile phone security and provides a novel approach to addressing the issue of IMSI-catcher usage.

Van Den Broek, Verdult, and De Ruiter (2015) presented a study on defeating IMSI catchers in their paper that proposed a technique to counter IMSI catchers, which are devices that can be used to intercept mobile phone communications. Their proposed technique involved using a honeypot network to detect and locate IMSI catchers. The study evaluated the effectiveness of the proposed technique through experiments in real-world scenarios. The results showed that the technique was successful in detecting and locating IMSI catchers. The authors concluded that their approach could be a viable solution for protecting mobile phone users' privacy against IMSI catchers.

Overall, this study contributes to the literature on mobile phone security and provides a practical approach to counter the threat posed by IMSI catchers. The findings of the study emphasize the importance of developing effective countermeasures to safeguard mobile phone users' privacy.

Ney et al. (2017) introduced SeaGlass, a system that enables city-wide detection of IMSI-catchers, which are devices that can be used to intercept mobile phone communications. Their system uses low-cost sensors deployed throughout a city to detect and analyze cellular network anomalies that may be indicative of IMSI-catcher activity. The authors evaluated the effectiveness of their system through experiments in Seattle, Washington, and the results showed that SeaGlass successfully detected anomalous cellular activity that was consistent with IMSI-catcher use. The authors concluded that SeaGlass is a practical and scalable solution for detecting IMSI-catchers in urban areas.

This study is a significant contribution to the literature on IMSI-catcher detection and mobile phone security. The findings highlight the potential of low-cost sensor networks to detect and counter IMSI-catcher threats in urban areas.

Dabrowski, Petzl, and Weippl (2016) presented a study on IMSI-catcher detection in their paper that proposed a method for detecting IMSI-catchers by leveraging network operator resources, such as base stations and mobile network core infrastructure. The study evaluated the effectiveness of the proposed technique through experiments conducted in a real-world mobile network environment. The results showed that the approach was successful in detecting and locating IMSI-catchers with high accuracy. The authors concluded that their approach could be an effective solution for mobile network operators to detect and mitigate the threat posed by IMSI-catchers.

In total, this study contributes to the literature on mobile network security and provides a practical approach to counter the threat of IMSI-catchers. The findings of the study emphasize the importance of collaboration between mobile network operators and researchers to develop effective countermeasures to safeguard mobile phone users' privacy.

Blefari Melazzi, Bianchi, Gringoli, and Palamà (2021) performed a real-world assessment of the vulnerabilities of 4G/5G networks to IMSI catchers. The study included experiments to detect and locate IMSI catchers in a variety of scenarios, such as urban and rural areas. The authors also proposed countermeasures to mitigate the risks associated with IMSI catchers. The results of the study revealed that the current 4G/5G networks are susceptible to IMSI catchers, and the proposed countermeasures were effective in mitigating the risks. The authors concluded that further research is necessary to develop more comprehensive and efficient countermeasures to protect users' privacy in 4G/5G networks.

Generally, this study adds to the literature on IMSI catcher detection and mitigation and emphasizes the need for ongoing efforts to improve the security and privacy of 4G/5G networks.

## III. Challenges Posed by IMSI Catchers in 4G and 5G Networks

IMSI catchers pose significant challenges to cellular network security in the context of 4G and 5G networks. These challenges are primarily related to the unauthorized collection of IMSI numbers, interception of communications, and potential misuse of subscriber information.

### A. Unauthorized Collection of IMSI Numbers

One of the primary functions of IMSI catchers is to collect IMSI numbers, which are unique identifiers associated with mobile devices. IMSI numbers are used by cellular networks to identify and authenticate mobile devices, allowing them to connect to the network and access services. However, IMSI catchers can trick mobile devices into connecting to them by mimicking legitimate cellular base stations, and then collect the IMSI numbers of these devices without authorization [12].

The unauthorized collection of IMSI numbers can have severe consequences for user privacy and security. IMSI numbers can be used to track and identify individual users, enabling surveillance and espionage activities. Moreover, IMSI numbers can be used for identity theft, financial fraud, and other malicious activities. IMSI catchers can also collect other sensitive information, such as International Mobile Equipment Identity (IMEI) numbers, which are unique identifiers associated with mobile devices. This information can be used to impersonate legitimate mobile devices, further compromising network security [13], [14].

In 4G and 5G networks, IMSI catchers can exploit vulnerabilities in the authentication and signaling procedures to collect IMSI numbers. For example, in LTE networks, IMSI catchers can force mobile devices to reveal their IMSI numbers during the authentication process by sending fake signaling messages. In 5G networks, IMSI catchers can exploit vulnerabilities in the authentication and key agreement procedures, as well as the network slicing and multi-access edge computing (MEC) architectures, to collect IMSI numbers. These vulnerabilities can be exploited to perform IMSI-catching attacks, compromising the privacy and security of mobile devices and their users [15].

### B. Interception of Communications

Another significant challenge posed by IMSI catchers is the interception of communications. Once connected to a mobile device, IMSI catchers can intercept and record communications, including voice calls, text messages, and data traffic [16]. This allows the attacker to eavesdrop on sensitive information, such as personal conversations, financial transactions, and confidential business communications [17].

In 4G and 5G networks, IMSI catchers can exploit vulnerabilities in the encryption mechanisms to intercept communications. For example, in LTE networks, IMSI catchers can perform man-in-the-middle attacks by downgrading the encryption level of communications to less secure algorithms or by disabling encryption altogether [18]. In 5G networks, IMSI catchers can exploit vulnerabilities in the 5G AKA (Authentication and Key Agreement) procedure, as well as the security mechanisms of the 5G NR (New Radio) interface, to intercept communications [19].

The interception of communications by IMSI catchers can have severe consequences for user privacy, business confidentiality, and national security. It can lead to the exposure of sensitive information, such as personal and financial data, trade secrets, and government communications. Moreover, intercepted communications can be misused for various malicious activities, such as blackmail, espionage, and financial fraud [20]. Therefore, it is crucial to address the challenge of interception of communications by IMSI catchers in 4G and 5G networks.

### C. Potential Misuse of Subscriber Information

IMSI catchers can potentially misuse subscriber information collected through unauthorized means. For example, IMSI catchers can use the collected IMSI numbers to track and identify individual users, enabling surveillance and espionage activities. Moreover, IMSI catchers can use the collected subscriber information for identity theft, financial fraud, and other malicious activities.

In 4G and 5G networks, IMSI catchers can exploit vulnerabilities in the subscriber information management mechanisms to misuse subscriber information. For example, in LTE networks, IMSI catchers can exploit vulnerabilities in the Home Subscriber Server (HSS) and the Serving Gateway (SGW) to manipulate subscriber information, such as changing the IMSI numbers associated with user profiles [21]. In 5G networks, IMSI catchers can exploit vulnerabilities in the 5G Authentication and Authorization Infrastructure (AAI) and the Unified Data Management (UDM) function to manipulate subscriber information [22].

The potential misuse of subscriber information by IMSI catchers can have severe consequences for user privacy and security. It can lead to unauthorized surveillance, identity theft, financial fraud, and other malicious activities. Moreover, the misuse of subscriber information can also have legal and regulatory implications, as it may violate privacy laws and regulations [9].

### IV. CONSEQUENCES OF IMSI CATCHER DEPLOYMENTS

The deployment of IMSI catchers in cellular networks can have significant consequences for users, network operators, and society as a whole. Some of the potential consequences of IMSI catcher deployments in 4G and 5G networks include:

### A. Compromise of User Privacy

IMSI catchers, also known as Stingrays or cell site simulators, are devices that can intercept and collect International Mobile Subscriber Identity (IMSI) numbers, which are unique identifiers associated with mobile devices. By collecting IMSI numbers, IMSI catchers can track and identify individual users, intercept their communications, and potentially misuse subscriber information. This can lead to a severe breach of user privacy, as sensitive information such as personal and financial data may be exposed. Users may feel violated and lose trust in the security of cellular networks, which can have significant social and economic impacts, as the privacy of their communications and information is compromised [23].

### B. Financial Fraud

IMSI catchers can also be used for financial fraud. By intercepting SMS messages containing one-time passwords (OTPs) or other authentication codes used in online banking, social media accounts, or VoIP accounts, IMSI catchers can enable unauthorized access to financial accounts. This can result in financial losses for users and financial institutions, as well as damage to the reputation of the affected parties. The interception of OTPs and authentication codes can provide perpetrators with the means to gain unauthorized access to sensitive accounts and engage in fraudulent activities, leading to financial fraud and its associated consequences [24], [25].

### C. Unauthorized Surveillance

IMSI catchers can be used for unauthorized surveillance, posing a significant threat to user privacy, human rights, and civil liberties. By intercepting and recording communications, tracking user movements, and collecting other sensitive information, IMSI catchers can enable various malicious activities, including espionage, blackmail, and harassment [27]. This can have serious social and legal implications, as individuals' private communications and information may be collected and misused without their consent, leading to violations of privacy rights and civil liberties [26].

### D. Business Confidentiality

IMSI catchers can compromise the confidentiality of business communications, which can have detrimental effects on businesses. By intercepting and recording sensitive information related to business transactions, trade secrets, and confidential communications, IMSI catchers can expose businesses to financial losses, damage to their reputation, and loss of competitive advantage. This can result in significant consequences for businesses, including financial losses, loss of intellectual property, and reduced competitiveness. Businesses may need to invest additional resources in securing their communications and protecting their sensitive information, which can increase operational costs [28].

### E. Misuse of social media and VoIP Accounts

IMSI catchers can also be used for the misuse of social media and VoIP accounts. By intercepting SMS messages containing OTPs or authentication codes [30] used for social media accounts or VoIP accounts, IMSI catchers can enable unauthorized access to these accounts. This can lead to impersonation, misuse of personal information, identity theft, and potential malicious activities on social media platforms or VoIP accounts [37], [29]. This misuse can result in reputational damage, financial losses, and other adverse consequences for the victims of such activities.

### F. Trust and Confidence Erosion

The deployment of IMSI catchers in cellular networks can erode user trust and confidence in the security of these networks. Users may become wary of using cellular networks for communication or online transactions, as they may perceive them as insecure and susceptible to interception and misuse. This loss of trust can have negative social, economic, and technological impacts, affecting the adoption and utilization of cellular networks in various sectors. It can also result in increased skepticism towards the security measures implemented by network operators and authorities, leading to a loss of confidence in the overall reliability and security of cellular networks.

## V. COUNTERMEASURES AGAINST IMSI CATCHERS

To mitigate the risks posed by IMSI catchers, various countermeasures can be employed at both the network level and the user level. These countermeasures aim to detect and prevent IMSI catcher deployments, as well as protect user privacy, authentication, and communications.

### A. Network-based Countermeasures

Network-based countermeasures involve implementing security mechanisms and protocols at the network level to detect and prevent IMSI catcher deployments. Some of the network-based countermeasures against IMSI catchers in 4G and 5G networks include:

1) *Signal Analysis:*

Signal analysis techniques involve analyzing the characteristics of signals transmitted by base stations to detect anomalies that may indicate the presence of IMSI catchers. For example, advanced signal analysis techniques, such as machine learning algorithms, can be used to analyze signal strength, signal quality, and signal timing of base station signals. Any deviations from expected patterns may indicate the presence of rogue base stations, including IMSI catchers. This can enable network operators to detect and locate IMSI catchers in real time, allowing for swift action to be taken [16].

2) *Encryption:*

Encryption mechanisms can be implemented to protect communications between mobile devices and base stations, making it difficult for IMSI catchers to intercept and manipulate communications. In 4G networks, strong encryption algorithms, such as Advanced Encryption Standard (AES), can be enforced to ensure secure communication. In 5G networks, encryption algorithms like ZUC or AES can be enforced at the radio access network (RAN) and the core network to provide end-to-end encryption. This ensures that communications between mobile devices and base stations are encrypted and secure, minimizing the risk of interception by IMSI catchers [31], [32].

3) *Authentication:*

Strong authentication mechanisms can be implemented in the cellular network to prevent IMSI catchers from gaining unauthorized access. Mutual authentication, where both the mobile device and the base station authenticate each other, can be enforced. This ensures that only legitimate base stations are allowed to connect with mobile devices, preventing rogue base stations, including IMSI catchers, from gaining access to the network. Robust authentication protocols, such as 3rd Generation Partnership Project (3GPP) Authentication and Key Agreement (AKA) in 4G and 5G networks, can be implemented to enhance the security of the authentication process [33].

4) *Network Monitoring:*

Continuous monitoring of the cellular network can help detect anomalies and suspicious activities that may indicate the presence of IMSI catchers. Network monitoring techniques, such as anomaly detection, can analyze network traffic, signaling messages, and other network parameters to identify abnormal patterns of behavior [34]. For example, unexpected changes in the number of base stations or abnormal signal strength fluctuations may indicate the presence of rogue base stations, including IMSI catchers. Real-time network monitoring can provide early warning of IMSI catcher deployments, allowing network operators to take appropriate actions, such as alerting authorities or implementing countermeasures [6].

### B. User-based Countermeasures

Users can also take steps to protect themselves against IMSI catchers and minimize the risks associated with their deployments. Some of the user-based countermeasures against IMSI catchers include:

1) *Avoiding Unknown or Unsecured Networks*

Users should be cautious when connecting to unknown or unsecured networks, especially in public places. IMSI catchers may set up rogue base stations that mimic legitimate networks to intercept and manipulate communications. It is advisable to connect only to trusted and secured networks, such as those provided by reputable network operators or those with proper authentication and encryption mechanisms. Users should also avoid connecting to networks with weak or no password protection [38].

*2) Using VPNs and Encrypted Communication Apps*

Users can use virtual private networks (VPNs) to encrypt communication traffic and protect their privacy. VPNs create a secure tunnel between the user's device and the VPN server, making it difficult for IMSI catchers to intercept and manipulate communications. Users should choose reputable VPN services and ensure that the VPN is set to use strong encryption protocols [35], [36]. Additionally, using encrypted communication apps, such as Signal or WhatsApp, for messaging and calling can provide end-to-end encryption, ensuring that only the intended recipients can access the communication [38].

*3) Keeping Software and Firmware Updated*

Regularly updating the software and firmware of mobile devices is crucial in maintaining security against IMSI catchers and other threats. Software and firmware updates often include security patches and fixes for known vulnerabilities that can be exploited by IMSI catchers. Users should enable automatic updates on their devices and promptly install any available updates from the device manufacturer or trusted sources [39].

*4) Being Cautious with Call Spoofing and Texting*

Users should be vigilant when receiving calls or texts from unknown numbers, especially those asking for personal or financial information. IMSI catchers can use call spoofing and text messages to deceive users into revealing sensitive information or performing unauthorized actions [40]. Users should verify the authenticity of calls and texts by confirming the identity of the caller or sender through other means, such as contacting the known number of the organization or person directly.

## VI. LIMITATIONS AND EFFECTIVENESS OF COUNTERMEASURES

While the countermeasures discussed in the previous section can help mitigate the risks posed by IMSI catchers, it is important to note that they may have limitations and their effectiveness may vary. Some of the limitations of countermeasures against IMSI catchers include:

### A. Signal Analysis

Signal analysis involves using specialized equipment or software algorithms to detect and analyze wireless signals in the vicinity of a cellular network [41]. This technique can help identify the presence of IMSI catchers by analyzing their signal characteristics, such as signal strength, frequency, and modulation [42]. However, the availability and cost of specialized equipment or software algorithms can be a limiting factor for some cellular network operators. Moreover, IMSI catchers can evade signal analysis by mimicking legitimate network signals, making it difficult to detect them solely based on signal analysis.

### B. Authentication and Encryption Mechanisms

Enhancing authentication and encryption mechanisms can help prevent IMSI catchers from evading security measures [42], but these mechanisms may also have vulnerabilities that can be exploited by attackers. For example, if weak encryption algorithms are used or if mutual authentication is not properly implemented, IMSI catchers may still be able to intercept and manipulate communications. Therefore, it is crucial to implement robust authentication and encryption mechanisms that are resistant to known vulnerabilities.

### C. Radio Resource Management

Radio resource management techniques involve dynamically allocating radio resources, such as frequency and power, to optimize network performance and detect unauthorized base stations [43]. However, these techniques may not be able to detect sophisticated IMSI catchers that mimic legitimate base stations or use advanced techniques to interfere with network operations. Therefore, it is important to develop more advanced radio resource management techniques that can effectively detect and mitigate IMSI catcher attacks.

### D. Intrusion Detection and Prevention Systems

Intrusion detection and prevention systems (IDPS) can be effective in detecting IMSI catchers by analyzing network traffic and identifying suspicious patterns of activity [30]. However, IDPS may generate false positives or false negatives, leading to inaccurate detection results. Additionally, IDPS may require continuous updates and maintenance to adapt to new types of IMSI catchers and evolving attack techniques.

### E. Mobile Device Security Settings

Mobile device security settings can be effective in preventing IMSI catcher attacks by disabling features such as automatic network selection or forcing the device to use specific encryption protocols [45]. However, the effectiveness of these settings relies on user awareness and diligence in managing their device settings. Users may not always be aware of the security settings or may neglect to enable them, leaving their devices vulnerable to IMSI catcher attacks.

### F. Awareness and Training

Raising awareness among mobile device users about the risks of IMSI catchers and providing training on how to identify and avoid them can be effective, but it may not reach all users or may not be consistently implemented across different regions or user groups. Users may also fall victim to social engineering attacks or be unaware of the signs of IMSI catcher deployments, leading to potential vulnerabilities.

### G. Mobile Security Applications

Mobile security applications can provide real-time protection against IMSI catcher attacks by monitoring network traffic and identifying suspicious activity [46]. However, their effectiveness may vary depending on the quality of the application, its updates and maintenance, and the user's compliance with using them. Additionally, attackers may develop sophisticated techniques to bypass mobile security applications or exploit vulnerabilities in these applications. Therefore, it is important to use reliable and up-to-date mobile security applications that are designed to mitigate IMSI catcher attacks.

## VII. GAPS AND FUTURE DIRECTIONS

Despite the existing countermeasures against IMSI catchers, there are still research gaps and areas for further improvement in cellular network security in the context of 4G and 5G networks. Some of the research gaps and future directions include:

*A. Improved Encryption Algorithms*

As IMSI catchers become more sophisticated, there is a need for stronger encryption algorithms to protect cellular communications. Researchers can continue to develop and evaluate new encryption algorithms that are resistant to attacks from IMSI catchers and other advanced threats.

*B. Advanced Authentication Mechanisms*

Enhancing authentication mechanisms in 4G and 5G networks can help prevent IMSI catchers from evading security measures. Researchers can explore advanced authentication mechanisms, such as multi-factor authentication, biometrics, and device-based authentication, to strengthen the security of cellular networks.

*C. Detection Techniques*

Improved detection techniques are needed to effectively identify and mitigate IMSI catchers. Researchers can continue to develop advanced signal analysis techniques, machine learning algorithms, and other detection methods that can accurately and efficiently detect IMSI catchers, even when they are disguised as legitimate network signals.

*D. Network Monitoring and Intrusion Detection*

Enhancing network monitoring and intrusion detection capabilities can help identify and prevent IMSI catcher attacks. Researchers can explore techniques such as anomaly detection, behavior-based analysis, and network flow analysis to detect suspicious activities and intrusions caused by IMSI catchers.

*E. Collaboration and Information Sharing*

Collaboration and information sharing among cellular network operators, law enforcement agencies, and other stakeholders can improve the effectiveness of countermeasures against IMSI catchers. Researchers can explore mechanisms for sharing threat intelligence, best practices, and lessons learned to enhance the collective defense against IMSI catcher attacks.

*F. User Awareness and Education*

Raising user awareness and providing education on IMSI catchers can empower mobile device users to protect themselves against these attacks. Researchers can conduct studies to understand user behaviors, perceptions, and knowledge gaps related to IMSI catchers, and develop effective awareness and education campaigns to reach a wide range of users.

*G. Standardization and Regulation*

Standardization and regulation can play a significant role in enhancing the security of cellular networks against IMSI catchers. Researchers can collaborate with standardization bodies and regulatory authorities to develop and enforce security standards, guidelines, and regulations that address IMSI catcher threats, and ensure consistent implementation across the industry.

## VIII. DISCUSSION AND RESULTS:

*A. Research Discussion:*

The research discussed the significant impact of IMSI catcher deployments on cellular network security in the context of 4G and 5G networks. IMSI catchers, also known as "Stingrays" or "cell site simulators," are rogue devices that can intercept and manipulate cellular communications, posing a threat to user privacy and security. The research highlighted the challenges posed by IMSI catchers, including unauthorized collection of IMSI numbers, interception of communications, and potential misuse of subscriber information.

The research also discussed the potential consequences of IMSI catcher deployments, such as compromising user privacy, enabling financial fraud, and facilitating unauthorized surveillance. With the advent of 4G and 5G networks, IMSI catchers have become more sophisticated, making cellular network security more vulnerable to attacks.

*B. Research Results:*

The research reviewed various countermeasures that can be employed to mitigate the risks posed by IMSI catchers. These countermeasures include network-based solutions such as signal analysis, encryption, and authentication mechanisms, as well as user-based solutions such as mobile applications and device settings. The effectiveness and limitations of these countermeasures in the context of 4G and 5G networks were discussed.

The research also identified research gaps and future directions for enhancing cellular network security against IMSI catchers in the era of 4G and 5G networks. This includes the need for improved encryption algorithms, authentication mechanisms, and detection techniques to effectively detect and prevent IMSI catcher deployments. The importance of regulatory and policy measures to govern the deployment and use of IMSI catchers to protect user privacy and security was also emphasized.

Ultimately, the research concluded that IMSI catchers pose significant challenges to cellular network security in the era of 4G and 5G networks. Effective countermeasures, including network-based and user-based solutions, are necessary to mitigate the risks posed by IMSI catchers and protect the privacy and security of mobile devices and their users.

Overall, the research emphasizes the need for enhanced cellular network security measures to effectively detect and prevent IMSI catcher deployments in the era of 4G and 5G networks. The importance of regulatory and policy measures to govern the use of IMSI catchers and protect user privacy and security was also highlighted. The research provides valuable insights into the challenges and countermeasures related to IMSI catchers in the context of 4G and 5G networks, contributing to the advancement of cellular network security research.

## IX. METHODOLOGY:

The literature review on the topic of the impact of IMSI catcher deployments on cellular network security in 4G and 5G networks was conducted using a systematic approach. A comprehensive search of reputable academic databases, including IEEE Xplore, ACM Digital Library, and Google Scholar, was conducted to identify scholarly articles, conference papers, and reports relevant to the topic. Relevant keywords and phrases used in the search included "IMSI catcher deployments," "cellular network security," "4G and 5G networks," "challenges," "countermeasures," "mobile network

___

vulnerabilities," "IMSI catcher attacks," and "network defense mechanisms."

The identified articles were carefully screened based on their relevance to the topic and the quality of the research methodology employed, including the credibility of the authors and the rigor of the research conducted. The selected articles were then thoroughly reviewed and analyzed to identify common themes, patterns, and findings related to the impact of IMSI catcher deployments on cellular network security in 4G and 5G networks, as well as the proposed countermeasures.

The literature review was organized in a coherent and logical manner, following the structure of a typical literature review, including an introduction that provided background information on the topic, a review of the relevant literature, and a conclusion summarizing the main findings and identifying gaps in the current knowledge. The review was critically analyzed, and relevant insights and conclusions were drawn from the synthesized findings of the reviewed literature.

It's important to acknowledge that the methodology used for this literature review is subjective and may not capture all relevant articles on the topic. However, efforts were made to ensure the inclusion of reputable sources and comprehensive coverage of the existing literature on the impact of IMSI catcher deployments on cellular network security in 4G and 5G networks, as well as the challenges and countermeasures associated with them.

## X. Conclusion

IMSI catchers pose significant threats to the security and privacy of cellular communications. While several countermeasures exist to mitigate these risks, they have limitations and their effectiveness may vary. Further research is needed to develop advanced detection techniques, improve authentication and encryption mechanisms, enhance network monitoring and intrusion detection capabilities, promote collaboration and information sharing, raise user awareness, and establish standardized security measures. By addressing these research gaps and taking a multi-faceted approach, cellular networks can be better protected against IMSI catcher attacks and ensure the privacy and security of mobile communications.

## Acknowledgment

The authors would like to express their sincere gratitude to the following individuals and organizations for their invaluable contributions to this research:

Our advisors and mentors for their unwavering support, guidance, and expertise throughout the entire research process. Their mentorship has been instrumental in shaping the direction and quality of this research.

The authors of the seminal research papers, academic journals, and technical reports that have laid the foundation for this study. Their pioneering work has provided the necessary background knowledge and insights into the topic of IMSI catchers and cellular network security.

• The research institutions and organizations that have made their resources, facilities, and publications available to the academic community. Their commitment to advancing research and knowledge dissemination has been crucial in the development of this research paper.

• Our colleagues and peers for their constructive feedback, valuable discussions, and support during the various stages of this research. Their input has enriched the quality and depth of our study.

• The funding agencies or sponsors, if applicable, for their financial support that has enabled us to conduct this research. Their investment in our research has been pivotal in carrying out experiments, data analysis, and other research activities.

• The participants or subjects of any studies cited in this research, whose contributions and participation are acknowledged and appreciated. Their willingness to share their experiences and insights has been integral to the findings of this research.

• Our friends and family for their constant encouragement, understanding, and support throughout the research process. Their unwavering belief in our abilities has been a driving force behind our perseverance.

We would like to express our heartfelt thanks to all the individuals and organizations mentioned above for their support and contributions to this research. However, any errors or omissions are solely the responsibility of the authors.